# Schema-based Scheduling of Event Processors and Buffer Minimization for Queries on Structured Data Streams*

Christoph Koch[†,‡]    Stefanie Scherzinger[§]    Nicole Schweikardt[♮]    Bernhard Stegmaier[♯]

‡: Technische Universität Wien, Vienna, Austria, Email: `koch@dbai.tuwien.ac.at`
§: Technische Universität Wien, Vienna, Austria, Email: `scherzinger@wit.tuwien.ac.at`
♮: Humboldt Universität zu Berlin, Berlin, Germany, Email: `schweikardt@informatik.hu-berlin.de`
♯: Technische Universität München, Munich, Germany, Email: `bernhard.stegmaier@in.tum.de`

## Abstract

We introduce an extension of the XQuery language, FluX, that supports event-based query processing and the conscious handling of main memory buffers. Purely event-based queries of this language can be executed on streaming XML data in a very direct way. We then develop an algorithm that allows to efficiently rewrite XQueries into the event-based FluX language. This algorithm uses order constraints from a DTD to schedule event handlers and to thus minimize the amount of buffering required for evaluating a query. We discuss the various technical aspects of query optimization and query evaluation within our framework. This is complemented with an experimental evaluation of our approach.

## 1 Introduction

XML is the preeminent data exchange format on the Internet. Stream processing naturally bears relevance in the data exchange context (e.g., in e-commerce). An increasingly important data management scenario is the processing of XQueries on streams of exchanged XML data. While the weaknesses of XML as a semistructured data model have been observed time and again (cf. e.g. [1]), XQuery on XML streams can be seen as the prototypical instance of the problem of queries on *structured* (vs. flat tuple) *data streams*.

Query engines for processing streams are naturally main-memory-based. Conversely, in some efforts towards developing main-memory XQuery engines whose original emphasis was *not* on stream processing (e.g., BEA's XQRL [9]), it was observed that it is worthwhile to build such systems using stream processing operators.

The often excessive need for buffers in current main memory query engines causes a scalability issue that has been identified as a significant research challenge [14, 13, 16, 7, 4]. While the efficient evaluation of XPath queries on streams has been worked on extensively in the past (here, state-of-the-art techniques use very little main memory) [2, 5, 10, 11], not much work has been done on efficiently processing XQuery on streams. The nature of XQuery, as a *data-transformation* query language entirely different from *node-selecting* XPath, requires new techniques for dealing with (and reducing) main memory buffers. State-of-the-art XQuery engines consume main memory in large multiples of the actual size of input XML documents [14].

Several recent projects have addressed XQuery on streams using transducer networks [13, 16]. Automata-based techniques are usually quite elegant but are hard to compare or integrate with other approaches and usually do not generalize to real-world query languages such as (full) XQuery with their great expressive power and all their odd features and artifacts of the standardization process. One approach [14] towards addressing the problem of reducing main memory consumption in an engine for full XQuery aims at reducing the amount of data buffered in main memory by pre-filtering the data read from the stream with the paths occurring in the query. However, for real-world XQueries, the need for substantial main memory buffers cannot be avoided in general.

An important goal is thus to devise a well-principled machinery for processing XQuery that is parsimonious with resources and allows to minimize the amount of buffering. Such machinery needs to be based on intermediate representations of queries that are syntactically close to XQuery and has to allow for an algebraic approach to query optimization, with buffering as an optimization target. This is necessary to allow for both extensibility and the leverage of a large body of related earlier work done by the database research community. However, to our knowledge, no principled work exists on query optimization in the framework of XQuery (rather than automata) for *structured data streams* (such as XML, but unlike flat tuple streams) which honors the special features of stream processing. Moreover, no framework for optimizing queries on structured data streams exists that captures the spirit of stream processing and allows for query optimization using schema information. (However, there are XQuery algebras meant for conventional query processing [18, 8], and there is work on applying them in the streaming context [7]. Moreover, the problem of optimizing XQueries using a set of constraints holding in the XML data model – rather than a schema – was addressed in [6].)

---
\* Extended version of [12].
† Work support by project Z29-N04 of the Austrian Science Fund (FWF).

In this paper, we attempt to improve on this situation. We introduce a query language, *FluX*, which extends XQuery by a new construct for event-based query processing called `process-stream`. FluX motivates a very direct mode of query evaluation on data streams (similar to query evaluation in XQRL [9]), and provides a strong intuition for what main memory buffers are needed in which queries. This allows for a strongly "buffer-conscious" mode of query optimization. The main focus of this paper is on automatically rewriting XQueries into event-based FluX queries and at the same time optimizing (reducing) the use of buffers using schema information from a DTD.

Consider the following XQuery $Q$ in a bibliography domain, taken from the XML Query Use Cases [19] (XMP Q3):

```
<results>
{ for $b in $ROOT/bib/book return
    <result> { $b/title } { $b/author } </result> }
</results>
```

For each book in the bibliography, this query lists its title(s) and authors, grouped inside a "result" element. Note that the XQuery language requires that, within each book, all titles are output before all authors.

The DTD

```
<!ELEMENT bib  (book)*>
<!ELEMENT book (title|author)*>
```

specifies that each `book` node may have several `title` and several `author` children. A priori, no order among these items is inferable from the given DTD. To implement this query, we may output the `title` children inside a `book` node as soon as they arrive on the stream. However, the output of the `author` children needs to be delayed (using a memory buffer) until we reach the closing tag of the `book` node (at that time, no further `title` nodes may be encountered). Then we may flush the buffer of `author` nodes, empty it, and later refill it with the `author` nodes from the next book.

We thus only need to buffer the `author` children of one `book` node at a time, but not the titles. Current main memory query engines do not exploit this fact, and rather buffer either the entire book nodes or, as an optimization [14], all `title` and all `author` nodes of `book`. Previous frameworks for evaluating or optimizing XQuery do not provide any means of making this seeming subtlety explicit and reasoning about it.

The `process-stream` construct of FluX allows to express precisely the mode of query execution just described. XQuery $Q$ is then phrased as a FluX query as follows:

```
<results>
{ process-stream $ROOT: on bib as $bib return
  { process-stream $bib: on book as $book return
      <result>
      { process-stream $book:
          on title as $t return {$t};
          on-first past(title,author) return
          { for $a in $book/author return {$a} } }
      </result> } }
</results>
```

A `process-stream $x` expression consists of a number of *handlers* which process the children of the XML tree node bound by variable `$x` from left to right. An "on a" handler fires on each child labeled "a" visited during such a traversal, executing the associated query expression. In the `process-stream $book` expression above, the `on-first past(title,author)` handler fires exactly once as soon as the DTD implies for the first time that no further `author` or `title` node can be encountered among the children of `$book`. (As observed above, in the given, very weak DTD, this is the case only as soon as the last child of `$book` has been seen.) In the query associated with the `on-first past(title,author)` handler, we may freely use paths of the form `$book/author` or `$book/title`, because such paths cannot be encountered anymore and we may assume that the query engine has already buffered all matches of these paths for us. It is a feasible task for the query engine to buffer only those paths that the query actually employs (see also [14]).

We call a query *safe* for a given DTD if, informally, it is guaranteed that XQuery subexpressions (such as the for-loop in the query above) do not refer to paths that may still be encountered in the stream. The above FluX query is safe: The for-expression employs the `$book/author` path, but is part of an on-first handler that cannot fire before all `author` nodes relative to `$book` have been seen.

If the path `$book/author` was replaced by, say, `$book/price` and the DTD production for `book` were

```
<!ELEMENT book ((title|author)*,price)>
```

then the FluX query above would not be safe. In that case, on the firing of `on-first past(title,author)`, the buffer for `$book/price` items would still be empty and the query result would be incorrect.

Query $Q$ can be processed more efficiently with the schema used in the XML Query Use Cases,

```
<!ELEMENT bib   (book)*>
<!ELEMENT book  (title,(author+|editor+),
                                publisher,price)>
```

Here, no buffering is required to execute our query because the DTD asserts that for each book, the title occurs strictly before the authors (we denote this as $Ord_{book}(\texttt{title}, \texttt{author})$, called an *order constraint*). We may phrase our query in FluX so as to directly copy titles and authors to the output as they arrive on the input stream. No data items need to be buffered.

```
<results>
{ process-stream $ROOT: on bib as $bib return
    { process-stream $bib: on book as $book return
        <result>
        { process-stream $book:
            on title  as $t return {$t};
            on author as $a return {$a} }
        </result> } }
</results>
```

The contributions of this paper are as follows.

- We introduce the FluX query language, which extends XQuery by the natural stream processing construct discussed above.

- We define the *safe* FluX queries (under a given DTD), which are those FluX queries in which XQuery subexpressions have the usual semantics (i.e., are not executed before the data items referred to have been fully read from the stream and may be assumed available in main memory buffers).

- We present an algorithm that schedules XQueries on streams using DTDs and transforms them into optimized FluX queries.

- We discuss the realization of query engines for FluX and the runtime buffer management.

- We have built a prototype FluX query engine which we evaluate by means of a number of experiments.

This is, to our knowledge, the first work on optimizing XQuery using schema constraints derived from DTDs[1]. A main strength of the approach taken in this paper is its extensibility, and even though space limitations require us to restrict our discussion to a (powerful) *fragment* of XQuery, our results can be generalized to even larger fragments. In our discussion at the end of the paper, we will also lay the foundations for algebraic optimization of queries using further information from the schema.

This paper is structured as follows. We start with basics on DTDs and regular languages in Section 2. Section 3 defines the query languages considered in this paper: Section 3.1 specifies an XQuery fragment. Based on this, Section 3.2 defines the FluX language, and Section 3.3 singles out the safe FluX queries. Section 4 presents our algorithms for translating XQuery into a particular normal form (Section 4.1) and for transforming this normal form into FluX (Section 4.2). Some examples of this transformation are given in Section 4.3. In Section 5 we discuss the implementation of our prototype system and the actual handling of buffers during query evaluation. In Section 6, we present our experiments, and we conclude with a discussion in Section 7.

## 2 Preliminaries

For simplicity of exposition, we consider the fragment of XML without attributes as our data model. Note that this is no substantial restriction, since attributes can be handled in the same way as subelements.

We focus on *valid* documents, i.e. documents conforming to a given *document type definition* (DTD).

Let $\Sigma$ be a set of symbols (or tag names). A DTD is an extended context free grammar over $\Sigma$. DTDs are *local* tree grammars [15], i.e. without competing nonterminals to the left-hand sides of productions, so each production in a DTD is unambiguously identified by a tag name in $\Sigma$.

Let $\rho$ be a regular expression and let $symb(\rho)$ be the set of atomic symbols that occur in $\rho$. By $L(\rho)$ we denote the language defined by $\rho$, i.e., the set of words over $symb(\rho)$ that are recognizable by $\rho$. Given a word $w$, let $w_i$ denote its $i$-th symbol. We define a binary relation $Ord_\rho \subseteq \Sigma \times \Sigma$ such that for $a, b \in \Sigma$,

$$Ord_\rho(a, b) \;:\Leftrightarrow\; \nexists w \in L(\rho) : w_i = b \wedge w_j = a \wedge i < j.$$

That is, $Ord_\rho(a, b)$ holds if there is no word in $L(\rho)$ in which a symbol $a$ is preceded by a symbol $b$. (All $a$ symbols occur before all $b$ symbols.) We refer to a constraint of the form $Ord_\rho(a, b)$ as an *order constraint*.

**Example 2.1** Let $\rho = (a^*.b.c^*.(d|e^*).a^*)$. Then, $Ord_\rho(b, c)$, $Ord_\rho(c, d)$, and $Ord_\rho(c, e)$, but $\neg Ord_\rho(a, c)$. $Ord_\rho$ is transitive, so we also have e.g. $Ord_\rho(b, d)$. □

DTDs have the nice property that regular expressions appearing in the right-hand sides of productions are *one-unambiguous*. This guarantees that an equivalent *deterministic* finite automaton can be computed in polynomial – even quadratic – time [3].

One can show the following:

**Proposition 2.2** *Given a regular expression $\rho$ from a DTD, $Ord_\rho$ can be computed in time $O(|\rho|^2)$.*

Let $\rho$ be a regular expression and let $S \subseteq \Sigma$. Then, for each word $u = u_1 \ldots u_n \in symb(\rho)^*$,

$$Past_{\rho,S}(u) \;:\Leftrightarrow\; \forall w \in symb(\rho)^* : uw \in L(\rho) \to \nexists\, i : w_i \in S,$$

$$first\text{-}past_{\rho,S}(u) \;:\Leftrightarrow\; Past_{\rho,S}(u) \,\wedge\, \big(n > 0 \to \neg Past_{\rho,S}(u_1 \ldots u_{n-1})\big).$$

Intuitively, when processing a word $uw \in L(\rho)$ from left to right, if $\mathit{first\text{-}past}_{\rho,S}(u)$ holds, then the reading of the last symbol of $u$ is the earliest possible time at which we know that none of the symbols in $S$ can be seen anywhere until the end of the word $uw$.

Appendix B shows how order constraints as well as *Past* and *first-past* events can be efficiently checked for a given DTD.

## 3 Query Language

In this section, we define the syntax and semantics of the FluX query language, which extends an XQuery fragment, denoted as XQuery$^{-}$, by a construct for event-based query processing.

Before defining FluX and XQuery$^{-}$, we need some more notation. We write $\$x$, $\$y$, $\$z$, ... to denote variables that range over XML trees. In the following, we overload the meaning of variable $\$x$ bound to an XML tree whose root is labeled $a$, by writing $\$x$ when we actually mean the DTD production unambiguously identified by the element $a$. For example, if the DTD contains the rule `<!ELEMENT a ` $\rho^a$`>` for a regular expression $\rho^a$, we write $Ord_{\$x}(c,d)$ instead of $Ord_{\rho^a}(c,d)$, and we write $symb(\$x)$ instead of $symb(\rho^a)$.

A *fixed path* is a sequence $a_1/\ldots/a_n$, where the $a_i$ are symbols from the DTD and $n \geq 1$. XPath expressions such as $a/*/b$, or $a//b$ or $a[b]$ are excluded.

An *atomic condition* is either of the form $\$x/\pi \; RelOp \; s$, `exists` $\$x/\pi$, or $\$x/\pi \; RelOp \; \$y/\pi'$, where $s$ is a string, $\pi$ and $\pi'$ are fixed paths, and $RelOp \in \{=, <, \leq, >, \geq\}$. A *condition* is a Boolean combination (using "and", "or", "not", and "true") of atomic conditions.

### 3.1 An XQuery Fragment: XQuery$^{-}$

**Definition 3.1 (XQuery$^{-}$)** The XQuery fragment XQuery$^{-}$ is the smallest set consisting of expressions

1. $\epsilon$ (the empty query)
2. $s$ (output of a fixed string)
3. $\alpha \; \beta$ (sequence)
4. `{ for` $\$x$ `in` $\$y/\pi$ `return` $\alpha$ `}` (for-loop)
5. `{ for` $\$x$ `in` $\$y/\pi$ `where` $\chi$ `return` $\alpha$ `}` (conditional for-loop)
6. `{` $\$x/\pi$ `}` (output of subtrees reachable from node $\$x$ through path $\pi$)
7. `{` $\$x$ `}` (output of subtree of node $\$x$)
8. `{ if` $\chi$ `then` $\alpha$ `}` (conditional)

where $\pi$ is a fixed path, $s$ a fixed string, $\chi$ a condition, and $\alpha$ and $\beta$ are XQuery$^{-}$ expressions.

Indeed, XQuery$^{-}$ is very similar to (a fragment of) standard XQuery [18], but differs in how we treat fixed strings inside queries. For example, the string `<hello>` is valid in XQuery$^{-}$, but not in standard XQuery. The query

```
<result> { $ROOT/bib/book } </result>
```

is understood in standard XQuery as a "result" node with an embedded query to produce its children. In the present paper, the same query is read as a sequence of three queries which write the string `<result>`, the /bib/book subtrees, and finally the string `</result>` to the output.

This, however, is only a subtlety which, on the one hand, is very convenient for obtaining our main results in Section 4 and which, on the other hand, as the following Proposition 3.2 shows, does not cause any problems. The alternative semantics of XQuery$^{-}$ is the basis of optimizations used *internally* by the query engine. Users formulate input queries in standard XQuery and may assume the usual semantics.

Let $[\![Q]\!]_{XQuery^{-}}(D)$ (resp., $[\![Q]\!]_{XQuery}(D)$) denote the XML document stream produced by evaluating query $Q$ on document $D$ under our XQuery$^{-}$ semantics (resp., under the standard XQuery semantics [18]).

**Proposition 3.2** *Let $Q$ be an XQuery that parses as an XQuery$^{-}$ query. Then, for any input document $D$, $[\![Q]\!]_{XQuery^{-}}(D) = [\![Q]\!]_{XQuery}(D)$.*

### 3.2 Syntax and Semantics of FluX

A *simple expression* is an XQuery$^{-}$ expression of the form $\alpha \; \beta \; \gamma$ where

- $\alpha$ and $\gamma$ are possibly empty sequences of strings and of expressions of the form "`{if` $\chi$ `then` $s$`}`", where $\chi$ is a condition and $s$ is a string.
- $\beta$ is either empty, "`{`$\$u$`}`", or "`{if` $\chi$ `then {`$\$u$`}}`", for some variable $\$u$ and some condition $\chi$.
- if $\beta$ is of the form "`{`$\$u$`}`", or "`{if` $\chi$ `then {`$\$u$`}}`", then no atomic condition that occurs in $\alpha \; \beta$ contains the variable $\$u$.

For instance,

```
<a>{$x}</a> {if $x/b=5 then <b>5</b>}
```

is a simple expression, but `{$x}{$y}` is not.

**Definition 3.3 (FluX)** The class of FluX expressions is the smallest set of expressions that are either *simple* or of the form

$$s \; \{ \texttt{process-stream} \; \$y: \; \zeta \} \; s'$$

where $s$ and $s'$ are possibly empty strings, $\$y$ is a variable, and $\zeta$ is a list (where entries are separated by semicolons ";") of one or more *event handlers*. Each event handler is of one of the following two types:

1. (so-called "**on-first**" handler)

    `on-first past(`$S$`) return` $\alpha$

    where $S \subseteq symb(\$y)$ and $\alpha$ is an XQuery$^{-}$ expression

2. (so-called "on" handler)

    on $a$ as $\$x$ return $Q$

where $\$x$ is a variable, $a$ is an element name in $symb(\$y)$, and $Q$ is a FluX expression.

We will use ps as a shortcut for `process-stream`, `on-first past(*)` as an abbreviation for `on-first past(symb($y))`, and furthermore `on-first past()` in place of `on-first past(∅)`.

Some examples of FluX expressions, as well as an informal description of the FluX semantics, were already given in Section 1; further examples can be found in Section 4.3. In general, we evaluate an expression

$$\{ \text{ process-stream } \$y: \ \zeta \ \}$$

as follows: An event-handling statement considers the children of the node currently bound by variable $\$y$ as a list (or stream) of nodes and processes this list one node at a time. On processing a node $v$ with children $t_1, \ldots, t_n$, with the labels of $t_i$ denoted as $label(t_i)$, we proceed as follows. For each $i$ from 0 to $n+1$ (i.e., $n+2$ times), we scan the list of event handlers $\zeta = \zeta_1; \ldots; \zeta_m$ once from the beginning to the end. In doing so, we test for each event handler $\zeta_j$ whether its event condition is satisfied, in which case the event handler $\zeta_j$ "fires" and the corresponding query expression is executed:

- A handler "on $a$ as $\$x$ return $Q$" fires if $1 \le i \le n$ and $label(t_i) = a$.
- A handler "on-first past($S$) return $\alpha$" fires if $0 \le i \le n$ and $first\text{-}past_{\$y,S}(label(t_1) \ldots label(t_i))$ is true (i.e., for the first time while processing the children of $\$y$, no symbol of $S$ can be encountered anymore) or if $i = n+1$ and this event handler has not fired in any of the previous $(n+1)$ scans.

In summary, it is well possible that several events fire for a single node, in which case they are processed in the order in which the handlers occur in $\zeta$. During the run on $t_1, \ldots, t_n$, each "on" handler may fire zero up to several times, while each "on-first" handler is executed exactly once.

For a FluX or XQuery$^-$ expression $Q$, let $free(Q)$ be the set of all *free variables* in $Q$, defined analogously to the free variables of a formula in first-order logic. That is, $free(\{\$x/\pi\}) = \{\$x\}$, and $free(\{\text{if } \chi \text{ then } \alpha\})$ consists of $free(\alpha)$ and the variables that appear in $\chi$. Further, $free(\{\text{for } \$x \text{ in } \$y/\pi \text{ return } \alpha\})$ contains the variable $\$y$ and the variables in $free(\alpha) \setminus \{\$x\}$. Finally, $free(\{\text{process-stream } \$y: \ \zeta \ \})$ consists of the variable $\$y$, and for each event handler in $\zeta$ of the form "on-first past($S$) return $\alpha$" also of the variables in $free(\alpha)$, and likewise for each event handler in $\zeta$ of the form "on $a$ as $\$x$ return $Q$" of the variables in $free(Q) \setminus \{\$x\}$.

Note that expressions of the form "{for $\$x$ in $\$y/a$ return $\alpha$}" and event handlers of the form "on $a$ as $\$x$ return $Q$" *bind* the variable $\$x$, i.e., remove it from the free variables of the superexpressions.

A FluX *query* is a FluX expression in which all free variables except for the special variable $\$ROOT$ corresponding to (the root of) the document are bound. That is, for a *query* $Q$ in FluX (resp. $\alpha$ in XQuery$^-$) we require that $free(Q) \subseteq \{\$ROOT\}$ (resp. $free(\alpha) \subseteq \{\$ROOT\}$).

As the following example shows, every XQuery$^-$ query can be transformed into a FluX query in a straightforward way.

**Example 3.4** Every XQuery$^-$ query $\alpha$ is equivalent to the FluX query

$$\{ \text{ ps } \$ROOT: \ \text{on-first past(*) return } \alpha \ \}$$

In Section 4 below we will show how, depending on a given DTD, this FluX query can be transformed into an equivalent FluX query that can be evaluated more efficiently. □

By the *size* of an expression $Q$, denoted $|Q|$, we refer to the size of its string representation.

By the *parent variable* of (FluX or XQuery$^-$) expression $\alpha$ in FluX query $Q$, denoted $parentVar(\alpha)$, we refer to the variable bound by the nearest superexpression of $\alpha$, or $\$ROOT$ if no such variable exists.

By the *condition paths* in $\alpha$, we refer to the set of paths $\$x/\pi$ in a condition $\chi$ that occurs in $\alpha$.

For FluX or XQuery$^-$ expressions $\alpha$ and $\beta$ we write $\alpha \preceq \beta$ (resp., $\alpha \prec \beta$) to denote that $\alpha$ is a subexpression (resp., proper subexpression) of $\beta$. An XQuery$^-$ subexpression $\alpha$ of a FluX expression $Q$ is called *maximal* if there is no XQuery$^-$ expression $\beta$ with $\alpha \prec \beta \preceq Q$. Note that a FluX query may contain several such maximal expressions.

**Example 3.5** The maximal XQuery$^-$ subexpressions of the first FluX query from Section 1 are "{$t}" and "{ for $\$a$ in $\$book/author$ return {$a} }". □

### 3.3 Safe Queries

We next define the notion of *safety* for FluX queries. Informally, a query is called *safe* for a given DTD if it is guaranteed that XQuery$^-$ subexpressions do not refer to paths that might still be encountered in an input stream compliant with the given DTD. For the precise definition we need the following notion.

The set of *dependencies* w.r.t. variable $\$y$ in a FluX or XQuery$^-$ expression $\alpha$ is defined as

$$dependencies(\$y, \alpha) :=$$
$$\{a \mid \text{ex. a condition path } \$y/a \text{ or } \$y/a/\pi \text{ in } \alpha\} \cup$$
$$\{b \mid \text{ex. } \$u, \pi, Q \text{ s.t. } \pi \text{ starts with symbol } b \text{ and }$$
$$\text{"{for } \$u \text{ in } \$y/\pi \text{ return } Q\text{)" } \preceq \alpha\}.$$

**Definition 3.6 (safe queries)** A FluX query $Q$ is called *safe* w.r.t. a given DTD if, and only if, for each subexpression "{ps $y$: $\zeta$ }" of $Q$, the following two conditions are satisfied:

1. For each handler "on-first past$(S)$ return $\alpha$" in the list $\zeta$, the following is true:

   - $\forall\, b \in dependencies(\$y, \alpha)$ we have: $b \in S$ or ex. $a \in S$ s.t. $Ord_{\$y}(b, a)$.

   - $\forall\, \$z \in free(\alpha)$ s.t. $\{\$z\} \preceq \alpha$ or $\{\$z/\pi\} \preceq \alpha$ (for some $\pi$) we have: $\$z = \$y$ and $\forall\, b \in symb(\$y)$: $b \in S$ or ex. $a \in S$ s.t. $Ord_{\$y}(b, a)$.

2. For each handler "on $a$ as $\$x$ return $\tilde{Q}$" in the list $\zeta$, and for each maximal XQuery$^-$ subexpression $\alpha$ of $\tilde{Q}$, the following is true:

   - $\forall\, b \in dependencies(\$y, \alpha)$ we have: $Ord_{\$y}(b, a)$
   - if $\alpha = \tilde{Q}$ (note that according to Definition 3.3 $\alpha$ must then be *simple*), then for all $\$u$ s.t. $\{\$u\} \preceq \alpha$ we have: $\$u = \$x$.

It can be shown that this notion of *safety* is sufficient to ensure that main memory buffers are fully populated when they are accessed by a query, i.e., that a FluX query can be evaluated in a straightforward way on input streams compliant with the given DTD.

Examples of safe FluX queries can be found in Sections 1 and 4. (To be precise, all FluX queries occurring in this paper are safe.)

## 4   Translating XQuery into FluX

In this section we address the problem of rewriting a query of our XQuery fragment into an equivalent FluX query that employs as little buffering as possible. This rewriting proceeds in two steps: First, we transform the given XQuery$^-$ query into an equivalent query in XQuery$^-$ *normal form* (Section 4.1). Afterwards, depending on a given DTD, this normalized query is rewritten into an equivalent safe FluX query (Section 4.2). The FluX extensions manage the event based, streaming execution of the query. All subqueries exclusively working on buffered data are XQuery$^-$ expressions.

### 4.1   A Normal Form for XQuery$^-$

An XQuery$^-$ expression is transformed into *normal form* by rewriting (subexpressions of) it using the rules in Figure 1 until no further changes are possible.

In an XQuery$^-$ expression in normal form, the following three properties hold: (1) All paths except those inside conditionals are simple-step paths, i.e. of the form $\$x/a$. (2) An expression in normal form does not contain any *conditional* for-loops, as the normalization process pushes conditionals inside the innermost for-loops. (3) For each subexpression of the form

$$\frac{\{\text{ for } \$x \text{ in } \$y/\pi \text{ where } \chi \text{ return } \beta \}}{\{\text{ for } \$x \text{ in } \$y/\pi \text{ return } \{ \text{ if } \chi \text{ then } \beta \} \}}$$

$$\frac{\{\ \$y/\pi\ \}}{\{\text{ for } \$x \text{ in } \$y/\pi \text{ return } \{\$x\} \}}$$

$$\frac{\{\text{ for } \$x \text{ in } \$y/a/\pi \text{ return } \beta \}}{\{\text{ for } \$x_0 \text{ in } \$y/a \text{ return }}{\{\text{ for } \$x \text{ in } \$x_0/\pi \text{ return } \beta \} \}} \quad (\$x_0 \text{ new})$$

$$\frac{\{\text{ if } \chi \text{ then } \{ \text{ for } \$x \text{ in } \$y/\pi \text{ return } \alpha \} \}}{\{\text{ for } \$x \text{ in } \$y/\pi \text{ return } \{ \text{ if } \chi \text{ then } \alpha \} \}}$$

$$\frac{\{\text{ if } \chi \text{ then } \alpha\ \beta \}}{\{\text{ if } \chi \text{ then } \alpha \}\ \{ \text{ if } \chi \text{ then } \beta \}}$$

$$\frac{\{\text{ if } \chi \text{ then } \{ \text{ if } \psi \text{ then } \alpha \} \}}{\{\text{ if } (\chi \text{ and } \psi) \text{ then } \alpha \}}$$

Figure 1: Normal form rewrite rules. Each rule is always applied downwards, i.e., the expression above the line is replaced by the expression below the line.

"{if $\chi$ then $\alpha$}", $\alpha$ is either a fixed string or of the form "{$\$x$}" for some variable $\$x$.

**Theorem 4.1** *The rule applications of Figure 1 can be implemented in such a way that the rewriting terminates for an input XQuery$^-$ expression $Q$ after $O(|Q|)$ rule applications with a unique result, the so-called* normalization *of $Q$, which is equivalent to $Q$.*

**Example 4.2 ([19], XMP, Q1)** Consider the following XQuery $Q_1$ for books published by Addison-Wesley after 1991, including their year and title.

```
<bib>
{ for $b in $ROOT/bib/book
      where $b/publisher = "Addison-Wesley" and
            $b/year > 1991
      return <book> {$b/year} {$b/title} </book> }
</bib>
```

We abbreviate the where-condition in the above query as $\chi$. Then $Q_1$ has the following normalization $Q_1'$:

```
<bib>
{ for $bib in $ROOT/bib return
    { for $b in $bib/book return
      { if $\chi$ then <book> }
      { for $year in $b/year return
        { if $\chi$ then {$year} } }
      { for $title in $b/title return
        { if $\chi$ then {$title} } }
      { if $\chi$ then </book> } } }
</bib>
```

□

### 4.2   Rewriting normalized XQuery$^-$ into FluX

To formulate our main rewrite algorithm for transforming normalized XQuery$^-$ queries into equivalent, safe FluX queries, we need some further notation.

```
1 function rewrite(Variable parentVar, Set⟨Σ⟩ H,
2                   XQuery⁻ β) returns FluXQuery
3 begin
4   let $x = parentVar;
5   if {$x} ⪯ β then
6   begin
7     if β is simple and dependencies($x, β) = ∅ then
8       return β
9     else
10      return { ps $x: on-first past(*) return β }
11  end
12  else /* {$x} ⋠ β */
13  begin
14    if β = β₁ β₂ then
15    begin
16      β₁' := rewrite(parentVar, H, β₁);
17      match ζ₁ such that β₁' = { ps $x: ζ₁ };
18      β₂' := rewrite(parentVar, H ∪ hsymb(ζ₁), β₂);
19      match ζ₂ such that β₂' = { ps $x: ζ₂ };
20      return { ps $x: ζ₁; ζ₂ }
21    end
22    else if β is simple then
23      /* β is either of the form s or { if χ then s } */
24      return { ps $x:
25               on-first past(dependencies($x, β) ∪ H)
26               return β }
27    else if β is of the form
28             { for $y in $z/a return α } then
29    begin
30      X := {b ∈ dependencies($x, α) ∪ H | ¬Ord_{$x}(b, a)};
31      if $z ≠ $x then
32        return { ps $x: on-first past(X) return β }
33      else if X ≠ ∅ then
34        return { ps $x: on-first past(X ∪ {a}) return β }
35      else
36      begin
37        α' := rewrite($y, ∅, α);
38        return { ps $x: on a as $y return α' }
39      end
40    end /* if β is for-expression */
41  end /* else {$x} ⋠ β */
42 end
```

Figure 2: Algorithm for rewriting XQuery⁻ into FluX.

Let $\Sigma$ be the set of tag names occurring in the given DTD. Let $\bot$ denote the empty list. For a list $\zeta$ of event handlers, we inductively define the set $hsymb(\zeta)$ of handler symbols for which an "on" handler or an "on-first" handler exists in $\zeta$:

$$hsymb(\bot) := \emptyset$$
$$hsymb(\zeta; \text{on } a \text{ as } \$x \text{ return } \alpha) := hsymb(\zeta) \cup \{a\}$$
$$hsymb(\zeta; \text{on-first past}(S) \text{ return } \alpha) := \\ hsymb(\zeta) \cup S$$

Our algorithm for recursively rewriting normalized XQuery⁻ expressions into FluX is shown in Figure 2. Note that this algorithm uses order constraints and hence depends on the underlying DTD. Given query $Q$, we obtain the corresponding FluX query as "rewrite($ROOT, \emptyset, Q$)". Some example runs of this al-gorithm are given in Section 4.3 below. The goals in the design of the algorithm were to produce a FluX query which (1) is safe w.r.t. the given DTD, (2) is equivalent to the input XQuery, and (3) minimizes the amount of buffering needed for evaluating the query in an XML document.

To meet goals (1) and (2), e.g. the particular order of the if-statements in the algorithm (lines 5, 14, 22, 27) is crucial. Also, a set $H$ of handler symbols must be passed on in recursive calls of the algorithm, because otherwise the resulting FluX query would not be safe. One important construct for meeting goal (3) is the case distinction in lines 31–39, where an "on" handler is created provided that this is *safe*, and an "on-first" handler is created otherwise.

**Theorem 4.3** *Given a DTD $D$ and a normalized XQuery⁻ query $Q$, "rewrite($ROOT, \emptyset, Q$)" runs in time $O(|D|^3 + |Q|^2)$ and produces a safe FluX query that is equivalent to $Q$ on all XML documents compliant with the given DTD.*

Our algorithm performs only a single traversal of the query tree. Runtime $O(|Q|^2)$ is mainly caused by the need to compute *dependencies*. Note that the resulting FluX query is in normal form.

## 4.3   Examples

We now discuss the effect of our rewrite algorithm on sample queries from the XQuery Use Cases [19].[2]

**Example 4.4 ([19], XMP, Q2)** Let us consider the XQuery $Q_2$ from the XQuery Use Cases [19], which creates a flat list of all the title–author pairs, with each pair enclosed in a **result** element. Due to space limitations we omit $Q_2$ here and only give its normalization $Q_2'$ (which is very similar to the original XQuery $Q_2$):

```
1 <results>
2 { for $bib in $ROOT/bib return
3   { for $b in $bib/book return
4     { for $t in $b/title return
5       { for $a in $b/author return
6         <result> {$t} {$a} </result> } } } }
7 </results>
```

When given a DTD that does not impose any order constraints on **title** and **author**, e.g., the first DTD from Section 1, then "rewrite($ROOT,\emptyset,Q_2'$)" proceeds as follows: First, $Q_2'$ is decomposed into two subexpressions $\beta_1$, consisting of line 1, and $\beta_2$, consisting of lines 2–7. Then, the rewrite algorithm is recursively called for $\beta_1$ and for $\beta_2$. As $\beta_1$ is *simple*, the call for $\beta_1$ produces the result

```
{ps $ROOT: on-first past() return <results> }
```

---

[2]We rewrite the queries to work without attributes.

The call for $\beta_2$ decomposes $\beta_2$ into two subexpressions $\beta_{21}$, consisting of lines 2–6, and $\beta_{22}$, consisting of line 7 of $Q_2'$. The recursive call "rewrite($\texttt{\$ROOT}, \emptyset, \beta_{21}$)" then executes lines 36–39 of the algorithm in Figure 2, because $\beta_{21}$ is a for-loop with parent variable $\texttt{\$ROOT}$ and associated set $X = X_{\beta_{21}} = \emptyset$. That is, the result

```
{ps $ROOT: on bib as $bib return α'₁ }
```

is produced, where $\alpha_1'$ is the result produced by the recursive call "rewrite($\texttt{\$bib}, \emptyset, \alpha_1$)", for the subquery $\alpha_1$ of $Q_2'$ in lines 3–6. This recursive call for $\alpha_1$ again executes lines 36–39 of the algorithm, producing the expression $\alpha_1' =$

```
{ps $bib: on book as $b return α'₂ }
```

where $\alpha_2'$ is the result of "rewrite($\texttt{\$b}, \emptyset, \alpha_2$)" for the subquery $\alpha_2$ of $Q_2'$ in lines 4–6. As $\alpha_2$ is a for-loop with parent variable $\texttt{\$b}$ and associated set $X = X_{\alpha_2} = \{\texttt{author}\}$, in this call line 34 of the algorithm is executed, producing the expression $\alpha_2' =$

```
{ps $b: on-first past(author,title) return α₂ }
```

All in all, "rewrite($\texttt{\$ROOT}, \emptyset, Q_2'$)" returns the following FluX query $F_2$:

```
1 {ps $ROOT:
2  on-first past() return <results>;
3  on bib as $bib return
4   {ps $bib: on book as $b return
5    {ps $b: on-first past(author,title) return
6     { for $t in $b/title return
7      { for $a in $b/author return
8       <result> {$t} {$a} </result> } } };
9  on-first past(bib) return </results> }
```

We will refer to the "{ps $\texttt{\$b}\cdots$}"-expression in lines 5–8 of $F_2$ as $\alpha_2'$. When evaluating the query $F_2$ on an XML document, the XQuery inside $\alpha_2'$ will be evaluated once *all* $\texttt{author}$ and *all* $\texttt{title}$ nodes have been encountered and buffered.

Let us now consider the case where we are given a DTD with the production

```
<!ELEMENT book (author*,title*)>
```

where the order constraint $Ord_{\texttt{book}}(\texttt{author}, \texttt{title})$ is met. While running "rewrite($\texttt{\$ROOT}, \emptyset, Q_2'$)" we now encounter the situation where $X = X_{\alpha_2} = \emptyset$ (rather than $\{\texttt{author}\}$, as with the previous DTD). Therefore, when processing the recursive call "rewrite($\texttt{\$b}, \emptyset, \alpha_2$)", now lines 36–39 of the algorithm are executed, eventually producing the following result $\alpha_2'' =$

```
{ps $b: on title as $t return
 {ps $t: on-first past(*) return
  { for $a in $b/author return
   <result> {$t} {$a} </result> } } }
```

Now, "rewrite($\texttt{\$ROOT}, \emptyset, Q_2'$)" yields query $F_2'$ differing from $F_2$ in the lines 5–8, which must be replaced by the above expression $\alpha_2''$.

When evaluating $F_2'$ on an XML document compliant with the second DTD, all $\texttt{author}$ nodes arrive before $\texttt{title}$ nodes and are buffered. Encountering a $\texttt{title}$ node in the input stream invokes the following actions: The value of that particular node is buffered, i.e., "$\texttt{on-first past(*)}$" delays the execution until the complete $\texttt{title}$ node has been seen. Then, we iterate over the buffer containing all collected $\texttt{author}$ nodes, each time writing the buffered $\texttt{title}$ and the current $\texttt{author}$ to the output. In contrast to the worst-case scenario above, we only buffer one title at a time in addition to the list of all authors. If there is more than one title, this strategy is clearly preferable. □

We next demonstrate that conditional for-loops are optimized correspondingly.

**Example 4.5 ([19], XMP, Q1)** Let us consider the query $Q_1$ and its normalization $Q_1'$ from Example 4.2. Given a DTD that does not impose any order constraints, e.g., the DTD

```
<!ELEMENT bib  (book)*>
<!ELEMENT book (title|publisher|year)*>
```

the function call "rewrite($\texttt{\$ROOT}, \emptyset, Q_1'$)" rewrites $Q_1'$ into the following FluX query $F_1$:

```
1  {ps $ROOT:
2   on-first past() return <bib>;
3   on bib as $bib return
4    {ps $bib: on book as $b return
5     {ps $b:
6      on-first past(publisher,year) return
7       { if χ then <book> };
8      on-first past(publisher,year) return
9       { for $year in $b/year return
10       { if χ then {$year} } };
11      on-first past(publisher,year,title) return
12       { for $title in $b/title return
13       { if χ then {$title} } };
14      on-first past(publisher,year,title) return
15       { if χ then </book> } } };
16  on-first past(bib) return </bib> }
```

The "$\texttt{on-first}$" handler in lines 11–13 delays query execution until all $\texttt{title}$ nodes have been buffered and all $\texttt{publisher}$ and $\texttt{year}$ nodes have been seen.

When given a different DTD, ensuring that both $Ord_{\texttt{book}}(\texttt{year}, \texttt{title})$ and $Ord_{\texttt{book}}(\texttt{publisher}, \texttt{title})$ hold, the $\texttt{title}$ nodes can be processed in the streaming fashion. The query $F_1'$ produced by "rewrite($\texttt{\$ROOT}, \emptyset, Q_1'$)" with this new DTD differs from the above query $F_1$ in the subexpression in lines 11–13 which must be replaced by

```
on title as $title return
 { if χ then {$title} }
```

Consequently, titles will not be buffered at all during evaluation of this query. □

Our rewrite algorithm is well capable of optimizing joins over two or more join predicates, as is demonstrated in the following example which is not part of the XQuery Use Cases.

**Example 4.6** We remain in the bibliography domain and consider documents compliant with the DTD

```
<!ELEMENT bib  (book|article)*>
<!ELEMENT book (title,(author+|editor+),publisher)>
<!ELEMENT article (title,author+,journal)>
```

The following XQuery $Q_3$ retrieves those authors of articles which are coauthored by people who have also edited books:

```
<results>
{ for $bib in $ROOT/bib return
  { for $article in $bib/article return
    { for $book in $bib/book
      where $article/author = $book/editor return
      { <result> {$article/author} </result> } }}}
</results>
```

For the remainder of this example, we abbreviate the join-condition comparing the authors of articles with the editors of books by $\chi$. Normalization yields the following query $Q_3'$:

```
1 <results>
2 { for $bib in $ROOT/bib return
3   { for $article in $bib/article return
4     { for $book in $bib/book return
5       { if χ then <result> }
6       { for $author in $article/author return
7         { if χ then {$author} } }
8       { if χ then </result> } } } }
9 </results>
```

When executing "rewrite($ROOT,∅,$Q_3'$)" with the DTD above, a recursive call "rewrite($bib,∅,$\beta$)" is eventually invoked for the subexpression $\beta$ of $Q_3'$ in lines 3–8. As $\beta$ is a for-loop with parent variable $bib and associated set $X = X_\beta = \{book\} \neq \emptyset$, line 34 of the algorithm is executed, returning an expression of the form `{ps $bib: on-first past(book,article) ⋯ }`. That is, as no order constraint between `article` and `book` holds, an `on-first` handler ensures that all articles and books will be buffered.

Altogether, "rewrite($ROOT,∅,$Q_3'$)" produces the following FluX query $F_3$, where $\alpha$ is used as abbreviation for the for-loop over books in lines 4–8 of $Q_3'$:

```
1 {ps $ROOT:
2   on-first past() return <results>;
3   on bib as $bib return
4     {ps $bib: on-first past(book,article) return
5       { for $article in $bib/article return α } };
6   on-first past(bib) return </results> }
```

When given a different DTD which imposes an order on books and articles, e.g. by the following production

```
<!ELEMENT bib (book*,article*)>
```

we can evaluate $Q_3'$ by buffering only `book` nodes but processing `article` nodes in a streaming fashion.

Indeed, when executing "rewrite($ROOT,∅,$Q_3'$)" with this new DTD, we eventually encounter the situation where set $X = X_\beta = \emptyset$, and therefore, lines 36–39 (rather than line 34, as with the previous DTD) are executed. Altogether, the FluX query $F_3'$ produced now, differs from the above query $F_3$ in the subexpression in lines 4–5, which must be replaced by

```
4 {ps $bib: on article as $article return
5   {ps $article: on-first past(author) return α }};
```

As all `book` nodes will have arrived before an `article` node can be encountered, data from books is available in buffers once the first `article` node is being read. When processing the children of an `article` node, we first buffer all `author` nodes before the query can be evaluated for the current article.

During the evaluation of $F_3'$, we therefore only buffer the authors of a single article in addition to the data already stored on books, whereas the evaluation of $F_3$ requires the authors of *all* articles to be buffered.  □

## 5 Implementation

In this section, we discuss our implementation of a query engine for evaluating FluX queries obtained from XQuery⁻ by the rewriting algorithm of the previous section.

We focus on the allocation of buffers and their use during query evaluation. Given a FluX query, we statically infer the buffers which are actually necessary in order to avoid superfluous buffering. Our prefiltering techniques generalize those of [14] to the scenario where certain parts of the input do not need to be buffered – even though they are used by the query – because they can be processed on-the-fly.

Buffers are implemented as lists of SAX events. The events stored in a buffer represent well-formed XML in the sense that start-element events and end-element events are properly nested within each other. This renders data read from (a stream replayed from) a buffer indistinguishable from data read from the input stream. In our implementation, we employ the same set of operators for handling both events originating from streams and from buffers.[3]

In the following, we say that a FluX query is in normal form if all of its (maximal) XQuery⁻ subexpressions are in normal form.

Let $Q$ be a safe FluX query in normal form. The FluX query engine identifies all nodes that must be stored in buffers, i.e. all nodes compared in join conditions, the roots of buffered subtrees that are output, and buffered nodes over which for-loops iterate.

---

[3]Thus, physical query evaluation proceeds in a way similar to that followed in XQRL [9].

More formally, let $\alpha$ be an XQuery$^-$ subexpression of $Q$. We define $\Pi(\$r, \alpha)$, the set of all buffered paths in $\alpha$ starting with variable $\$r$, as $\Pi(\$r, \epsilon) = \Pi(\$r, s) = \emptyset$, $\Pi(\$r, \{\$r\}) = \{\$r\}$, $\Pi(\$r, \alpha\beta) = \Pi(\$r, \alpha) \cup \Pi(\$r, \beta)$,

$$\Pi(\$r, \{\texttt{for } \$x \texttt{ in } \$y/a \texttt{ return } \alpha\}) =$$
$$\Pi(\$r, \alpha) \cup \{\$r/a \mid \$y = \$r \text{ and } \Pi(\$x, \alpha) = \emptyset\}$$
$$\cup \{\$r/a/w \mid \$y = \$r \text{ and } \$x/w \in \Pi(\$x, \alpha)\},$$

and $\Pi(\$r, \{\texttt{if } \chi \texttt{ then } \alpha\}) = \Pi(\$r, \alpha) \cup \{\$r/\pi \mid \chi$ contains an atomic condition $\$r/\pi \, RelOp \, \$y/\pi'$ or $\$y/\pi' \, RelOp \, \$r/\pi\}$.

For a variable $\$r$ and a safe FluX query $Q$ in normal form, we now define $\Pi(\$r)$ as the union of all $\Pi(\$r, \alpha)$ s.t. $\alpha$ is a maximal XQuery$^-$ subexpression of $Q$.

Let $\mathcal{T}(\$r)$ be the prefix tree constructed by merging all paths from $\Pi(\$r)$. Intuitively, the prefix tree defines a projection of the input document, as it describes which parts of the input tree will be buffered.

We optimize the prefix tree in order to restrict the amount of data being buffered. Let $\mathcal{T}^m(\$r)$ be the tree obtained from $\mathcal{T}(\$r)$ by marking each node $v$ if $v$ either occurs in a join condition or the entire subtree rooted at node $v$ is output and must therefore be buffered. For unmarked nodes in $\mathcal{T}^m(\$r)$, we merely store the SAX events for the opening and the closing tag.

Clearly, if a node is marked and we buffer it together with its subtree, we also buffer the subtrees of any descendant nodes at the same time. Thus we only buffer the data of the topmost marked nodes in $\mathcal{T}^m(\$r)$. For example, if we need to buffer two subtrees reachable by paths $\pi$ and $\pi'$ respectively, where $\pi$ is a prefix of $\pi'$, we restrict ourselves to buffering the subtree identified by $\pi$. Let $\mathcal{T}^p(\$r)$ be the pruned prefix tree obtained from $\mathcal{T}^m(\$r)$ by successively removing a subtree rooted at node $v'$ if an ancestor node $v$ is marked. We refer to $\mathcal{T}^p(\$r)$ as a *buffer tree*.

W.l.o.g., below we will assume that variables in queries are used uniquely, i.e., each variable name is bound at most at one place in the query. For a safe FluX query $Q$ in normal form, let $X$ be the set of variables that are free in maximal XQuery$^-$ subexpressions of $Q$. The variables in $X$ are precisely those for which we will later define buffers.

**Example 5.1** The following FluX query selects all book publishers whose CEO has published articles.

```
{ ps $ROOT: on bib as $bib return
 { ps $bib: on article as $article return
  { ps $article: on-first past(author) return
    { for $book in $bib/book return
     { for $p in $book/publisher return
       { if $article/author = $book/publisher/ceo
          then {$p} } } } } }
```

Here, $X = \{\$\texttt{bib}, \$\texttt{article}\}$ and we compute the sets

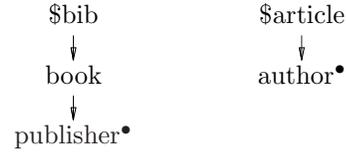

$bib $article

book author•

publisher•

Figure 3: Buffer trees of variables $bib and $article.

of buffer paths for each variable in $X$:

$$\Pi(\$\texttt{bib}) = \{\$\texttt{bib/book/publisher/ceo},$$
$$\$\texttt{bib/book/publisher}\}$$
$$\Pi(\$\texttt{article}) = \{\$\texttt{article/author}\}$$

We construct a buffer tree for each variable in $X$ with a nonempty set of buffer paths. Here, we obtain the trees shown in Figure 3 (the bullet denotes a marked node). Note that the leaf node `ceo` has been pruned off the buffer tree of variable `$bib`. □

We evaluate a safe FluX query $Q$ in normal form as follows. We compute $X$ and construct the buffer tree $\mathcal{T}^p(\$r, Q)$ for each variable $\$r \in X$.

We further associate an evaluator function with each variable $\$r$ in $X$. When variable $\$r$ is bound to a node $v$ of the incoming XML tree, then the evaluator `Eval_$r` is responsible for handling all events generated while processing the children of $v$.

We define a buffer `buffer_$r` for variable $\$r$ in the evaluator which calls `Eval_$r`. This buffer is initialized on entering the scope of $\$r$ and freed on re-entering it. The buffer tree of $\$r$ can be considered a schema for all events stored in buffer `buffer_$r`. At the same time, the buffer tree determines how the set of evaluators is to be extended such that all buffers are correctly filled.

For a node $v$ in buffer tree $\mathcal{T}^p(\$r)$, reachable by path $\$r/a_1/\ldots/a_n$, we implement the corresponding buffering strategy in a set of evaluators. Starting with `Eval_$r` and $a_1$, we successively extend the evaluators responsible for handling the children of a node labeled $a_i$ such that the events for the opening and closing tag of the respective node will be added to `buffer_$r`. In cases where no such evaluator exists, we introduce new variables and evaluators accordingly. As there is at most one case statement in an evaluator for an "on a" event under a given node label "a", it is clear where corresponding commands are to be introduced. In case $a_n$ is marked, we insert the respective code for adding all events corresponding to node $a_n$ and its subtree to `buffer_$r`.

**Example 5.2** Consider query $F_3'$ of Example 4.6. The set of buffer trees for this query is as in Figure 3 with the publisher node replaced by an editor node.

Its FluX evaluation strategy is as follows.

```
for each event e of Eval_$ROOT, switch(e) {
    case(ofp()):    { output "<results>"; }
    case(on bib):   { initialize(buffer_$bib);
                      Eval_$bib(node(e)); }
    case(ofp(bib)): { output "</results>"; } }

for each event of Eval_$bib, switch(e) {
    case(on book):    { buffer_$bib.add(<book>);
                        Eval_$book(node(e));
                        buffer_$bib.add(</book>); }
    case(on article): { initialize(buffer_$article)
                        Eval_$article(node(e)); } }

for each event of Eval_$book, switch(e) {
    case(on editor):  { buffer_$bib.add(node(e));} }

for each event of Eval_$article, switch(e) {
    case(on author):
            { buffer_$article.add(node(e)); }
    case(ofp(author)):  { execute_subquery(α); } }
```

At the beginning of the stream, the evaluator of variable `$ROOT` handles the event `on-first past()`, denoted `ofp()` above, by writing the opening tag `<results>` to the output. Correspondingly, the event `ofp(bib)` signals the end of the stream and the closing tag `</results>` is output.

Yet when processing the SAX event for the opening tag of root node `bib`, the buffer associated with variable `$bib` is initialized and the evaluator for `$bib` takes over to handle all events generated while parsing the children of `bib`.

Book nodes arriving on the stream are stored in the buffer of variable `$bib`, while `editor` nodes are buffered by the evaluator for variable `$book` together with their complete subtrees. As with all safe FluX queries, we may rely on the fact that buffer `buffer_$bib` is filled by the time we encounter the first article node. The buffer is not freed until the complete subtree under the node bound to variable `$bib` has been parsed.

Processing the children of an `article` node, any authors are first buffered in `buffer_$article` until the event `ofp(author)` guarantees that the subquery $\alpha$ of $F_3'$ can be executed correctly.    □

Join conditions are handled similarly, by buffering both constituent paths of the condition. Simple conditions comparing a path with a constant can be evaluated on the fly while reading the paths, so only a Boolean flag is required, which has to be appropriately initialized upon entering the relevant variable scope.

## 6  Experiments

In order to assess the merits of the approach presented in this paper, we have experimentally evaluated our prototype query engine implemented in JAVA using a number of queries on data obtained using the XMark benchmark generator.

|       | | FluX | Galax | AnonX |
|-------|------|------|-------|-------|
| $Q_1$ | 5M   | 2.1s/0 | 13.4s/37M | 3.4s |
|       | 10M  | 2.8s/0 | 29.8s/83M | 6.7s |
|       | 50M  | 7.8s/0 | - / >500M | 38.3s |
|       | 100M | 14.0s/0 | - / >500M | - |
| $Q_8$ | 5M   | 6.8s/1.54M | 296.9s/50M | 143.8s |
|       | 10M  | 17.2s/3.16M | 1498.3s/100M | 534.8s |
|       | 50M  | 357.8s/16.00M | - / >500M | - |
|       | 100M | 11566.9s/32.25M | - / >500M | - |
| $Q_{11}$ | 5M   | 5.6s/374k | 277.0s/50M | n/a |
|       | 10M  | 11.4s/741k | 1663.7s/100M | n/a |
|       | 50M  | 170.8s/3.64M | - / >500M | n/a |
|       | 100M | 626.8s/7.27M | - / >500M | n/a |
| $Q_{13}$ | 5M   | 2.2s/0 | 12.8s/38M | 3.0s |
|       | 10M  | 3.1s/0 | 27.2s/73M | 5.2s |
|       | 50M  | 7.9s/0 | 230.1s/344M | 88.0s |
|       | 100M | 13.9s/0 | - / >500M | - |
| $Q_{20}$ | 5M   | 2.8s/4.66k | 13.2s/36M | 2.5s |
|       | 10M  | 3.4s/5.18k | 29.7s/80M | 6.2s |
|       | 50M  | 8.7s/7.01k | - / >500M | 151.9s |
|       | 100M | 15.4s/7.02k | - / >500M | - |

Figure 4: Benchmark results.

Our implementation supports the XQuery$^-$ fragment as defined in Section 3. We took selected queries of the XMark benchmark and, as XQuery$^-$ does not include certain features that are used in these queries, adapted them correspondingly. In detail, attributes were converted into subelements of their parent element in our tests (the XMark DTD was adjusted accordingly). Occurrences of the XPath step `text()` were replaced by {$x} expressions that print the whole element instead. We eliminated the `count($x)` aggregations by again outputting $x$ instead. XMark queries 1, 8, 11, and 13 were adjusted as sketched above. We extracted the last FLWR subexpression of original query 20 (which computes persons whose income is not available) for our novel query 20. The queries thus obtained can be found in full in Appendix A.

We used data generated by the XMark xmlgen data generation tool (V. 0.96) of the sizes 5MB, 10MB, 50MB, and 100MB as input data. All tests were performed with the SUN JDK 1.4.2_03 and the built-in SAX parser on an AMD Athlon XP 2000+ (1.67GHz) with 512MB RAM running Linux (gentoo linux using kernel 2.6). Our query engine was implemented precisely as described in this paper. As a reference implementation the Galax query engine (V. 0.3.1) was employed with projection turned on [14]. The performance of query evaluation was studied by measuring the execution time[4] (in seconds) and maximum memory consumption (in bytes) of each engine. The memory and CPU usage of both query engines were mea-sured by internal monitoring functions (excluding the fixed memory consumption of the Java Virtual Machine).

---

[4]The times taken for query rewriting were negligible and are not reported separately in our experiments.

To give a broader overview over the performance of our approach we evaluated our queries additionally with a commercial XQuery system of a major company that has to remain anonymous and will be called AnonX below. Unfortunately, we could not determine the exact memory consumption for this system. Hence, we only state its execution time. As AnonX was not able to parse Query 11, we are not able to list the execution time.

Figure 4 shows the results of our experiments. To evaluate most queries with input greater than 10MB, Galax needed more than 500MB of main memory after running for a few minutes (which caused the system to start swapping). These runs were aborted. Obviously, our prototype engine clearly outperforms Galax with respect to both execution time and memory consumption. Queries 1 and 13 are evaluated on-the-fly without any buffering because of the order constraints imposed by the DTD. Query 20 has to buffer only a single element at a time, which leads to very low memory consumption in comparison to the traditional approach. Queries 8 and 11 perform a join on two subtrees (i.e. of `people` and `closed_auction` resp. `open_auction`) and therefore inevitably have to buffer elements. Nevertheless, due to our effective projection scheme only a small fraction of the original data is buffered. The rapid increase in execution time is due to the fact that we compute joins by naive nested loops at the moment. (We will work on this orthogonal but vital issue in the future.)

The comparison of the execution times to AnonX again shows the competitiveness of our query engine. AnonX ran out of memory processing queries marked by "-" (the maximum heap size of the Java VM was set to 512MB in both cases) and hence did not give any results in this case.

Altogether, our optimization approach seems to perform very well with respect to execution time, maximum memory consumption, and the maximum size of XML documents that can be processed.

## 7 Discussion and Conclusions

Main memory is probably the most critical resource in (streamed) query processing. Keeping main memory consumption low is vital to scalability and has – indirectly – a great impact on query engine performance in terms of running time.

The main contribution of this paper is the FluX language together with an algorithm for automatically translating a significant fragment of XQuery into equivalent FluX queries. FluX – while intended as an internal representation format for queries rather than a language for end-users – provides a strong intuition for buffer-conscious query processing on structured data streams. The algorithm uses schema information to schedule FluX queries so as to reduce the use of buffers.

As evidenced by our experiments, our approach indeed dramatically increases the scalability of main memory XQuery engines, even though we think we are not yet close to exhausting this approach, neither with respect to run-time buffer management and query processing nor query optimization.

In particular, further constraints such as cardinality constraints derived from the DTD, telling, e.g., that a `book` node has at most one `publisher` child (let this be denoted by $\texttt{publisher} \in \|\|_{\texttt{book}}^{\leq 1}$), could be used to simplify XQueries before they are rewritten into FluX using rewrite rules such as

$$\frac{\{\ \texttt{for}\ \$x\ \texttt{in}\ \$r/a\ \texttt{return}\ \alpha\ \}}{\{\ \texttt{for}\ \$x\ \texttt{in}\ \$r/a\ \texttt{return}\ \beta\ \}} \quad \left(a \in \|\|_{\$r}^{\leq 1}\right)$$

which can form the basis of algebraic query optimization for buffer minimization.

Sequences of for-loops iterating over singletons are a natural product of the normalization process that we have described. For example, the query

```
{ for $b in $ROOT/book return
    {$b/publisher/name} {$b/publisher/address} }
```

uses a sequence of two loops over `publisher` in its normal form, which can be rewritten into one using the rule above. By first merging for-loops, it is often possible to obtain FluX queries that require no buffering at all, while two subsequent loops over the same path generally cause that path to be buffered.

Another important optimization is to push `if`-expressions – which we have moved down the query tree to obtain our normal form – back "up" the expression tree as soon as the other simplifications have been realized.

# APPENDIX

In Appendix A we list the XQueries used as benchmark queries in Section 6. Appendix B explains how *order constraints* as well as *Past* and *first-past* events can be efficiently checked for a given DTD.

## A   Benchmark Queries

In this section we present the queries used in our benchmark experiments. As we have briefly sketched in Section 6, we have adapted selected queries of the XMark benchmark to suit the capabilities of our prototype implementation. In detail, we made the following changes:

- Attributes were converted into subelements of their parent element. For example,

```
<person id = "..."> ... </person>
```

was converted to

```
<person>
  <person_id>
    ...
  </person_id>
  ...
</person>
```

The XMark queries and the schema were adapted accordingly. While processing an XML stream (generated by the XMark xmlgen data generation tool), our XSAX parser converted attributes into subelements on-the-fly.

- XPath steps like `text()` were omitted; the whole element was printed instead.

- Aggregations such as `count($x)` were omitted and (a subtree of) element $x$ was written to the output instead.

The current version of our prototype implementation already supports an XQuery fragment that is slightly larger than the class $XQuery^-$ defined in the present paper. For example, in the outermost for-loops of queries, the variable `$ROOT` may be omitted, and where-conditions may also use statements such as $x/\pi > c * y/\pi'$, for constants $c$ (cf., Query 11 below) and `empty($x/\pi)` (cf., Query 20 below). Note the latter condition is equivalent to the $XQuery^-$ condition `not exists` $x/\pi$.

The following queries were used in our experiments (for all systems):

### Query 1

```
<query1>
{ for $b in /site/people/person
  where $b/person_id = 'person0'
  return
    <result> {$b/name} </result> }
</query1>
```

### Query 8

```
<query8>
{ for $p in /site/people/person return
  <item>
    <person> {$p/name} </person>
    <items_bought>
      { for $t in
```

```
              /site/closed_auctions/closed_auction
          where $t/buyer/buyer_person = $p/person_id
          return
              <result> {$t} </result> }
      </items_bought>
    </item> }
</query8>
```

### Query 11

```
<query11>
  { for $p in /site/people/person return
    <items>
      {$p/name}
      { for $o in /site/open_auctions/open_auction
        where $p/profile/profile_income >
                              (5000 * $o/initial)
        return
          {$o/open_auction_id} }
    </items> }
</query11>
```

### Query 13

```
<query13>
  { for $i in /site/regions/australia/item return
    <item>
      <name> {$i/name} </name>
      <desc> {$i/description} </desc>
    </item> }
</query13>
```

### Query 20

```
<query20>
  { for $p in /site/people/person
    where empty($p/person_income)
    return $p }
</query20>
```

## B  Efficient Checking of Schema Constraints

By a *marking* of a regular expression $\rho$ we denote a regular expression $\rho'$ such that each occurrence of an atomic symbol in $\rho$ is replaced by the symbol with its position among the atomic symbols of $\rho$ added as a subscript. That is, the $i$-th occurrence of a symbol $a \in symb(\rho)$ is replaced by $a_i$. The reverse of a marking (indicated by #) is obtained by dropping the subscripts.

All regular expressions in a DTD are one-unambiguous [3]. Intuitively, a one-unambiguous regular expression $\rho$ allows for deterministic matching of a word $w \in L(\rho)$ using only a one-token lookahead.

For each one-unambiguous regular expression, an equivalent deterministic finite automaton called the Glushkov automaton [3] can be constructed efficiently – in *quadratic time*. Glushkov automata have the characteristic properties that (1) each state in a Glushkov automaton (apart from the initial state) corresponds to a symbol in the marked regular expression, and (2) each transition $\delta(q, a) = p$ into a state $p$ takes place under input symbol $a = p^{\#}$. We refer to [3] for a formal definition of one-unambiguous regular expressions, of Glushkov automata, and their construction.

Let $\mathcal{G} = (Q, symb(\rho), \delta, q_0, F)$ be a Glushkov automaton for $\rho$ and let $\delta^*(q, a_1 \ldots a_n) = \delta(\ldots \delta(\delta(q, a_1), a_2), \ldots, a_n)$. Let $\Delta$ be the reachability relation

$$\{\langle q_i, q_j \rangle \mid \exists u \in symb(\rho)^* : \delta^*(q_i, u) = q_j\}.$$

Obviously, $\Delta$ can be computed in time $O(|Q|^2)$ by simply, for each $q \in Q$, computing the reachable states in the transition graph of $\mathcal{G}$ (for which there is a well-known linear time algorithm [17]).

We define the relation $Past_\rho \subseteq Q \times \Sigma$ as

$$Past_\rho(q_i, a) \Leftrightarrow \nexists q_j : q_j^{\#} = a \wedge \langle q_i, q_j \rangle \in \Delta.$$

Intuitively, $Past_\rho(q_i, a)$ means that on reaching state $q_i$, we are past all occurrences of $a$ (i.e., we may not encounter $a$ anymore until we reach the end of the word, otherwise it is not in $L(\rho)$.) Obviously, this relation can be obtained from $\Delta$ in time $O(|Q|^2)$. (Note that # is functional and imposes a partition of $Q$.)

Now we can define order constraints as

$$Ord_\rho(a, b) \Leftrightarrow \forall q : (q^{\#} = b) \rightarrow Past_\rho(q, a).$$

It is easy to see that the relation $Ord_\rho$ can be computed in time $O(|Q| \cdot |symb(\rho)|)$.

Given a set $S \subseteq \Sigma$, we can pre-compute a table

$$PastTable_{\rho, S}(q) \Leftrightarrow \forall a \in S : Past_\rho(q, a).$$

We assume that the input XML stream is processed by a SAX parser, which validates each token read from the input by simulating a transition of the Glushkov automaton associated with the current DTD production. In each such transition, we may compute *first-past*$_{\rho, S}$ by a constant-time lookup in $PastTable$.

More precisely, we compute *first-past*$_{\rho, S}(u_1 \ldots u_n)$ as follows. Initially,

$$\textit{first-past}_{\rho, S}(\epsilon) := PastTable_{\rho, S}(q_0).$$

On making the transition $\delta(q, u_n) = q'$ where $q = \delta^*(q_0, u_1 \ldots u_{n-1})$,

$$\textit{first-past}_{\rho, S}(u_1 \ldots u_n) :=$$
$$PastTable_{\rho, S}(\delta(q, u_n)) \wedge \neg PastTable_{\rho, S}(q).$$

Thus the SAX parser generates on-first-past punctuation events (which fire when *first-past*$_{\rho, S}$ becomes true) in addition to traditional SAX events with very little overhead, namely one validating DFA transition and one constant-time lookup per input token read.